\documentclass[letterpaper]{JHEP}
\usepackage{epsfig}
\usepackage{graphicx}
\usepackage{amssymb,amsmath}
\usepackage{epstopdf}

\def\be{\begin{eqnarray}}
\def\ee{\end{eqnarray}}

\title{Holographic $U(1)_A$ and String Creation}

\author{Oren Bergman\\
Department of Physics\\
Technion, Israel Institute of Technology\\
Haifa 32000, Israel\\
\email{bergman@physics.technion.ac.il}}

\author{Gilad Lifschytz\\
Department of Mathematics and Physics and CCMSC \\
university of Haifa at Oranim\\
Tivon 36006, Israel \\
\email{giladl@research.haifa.ac.il}}

\date{}

\abstract{We analyze the resolution of the $U(1)_A$ problem in the Sakai-Sugimoto 
holographic dual of large $N_c$ QCD at finite temperature. It has been shown that
in the confining phase the axial symmetry is broken at order $1/N_c$,
in agreement with the ideas of Witten and Veneziano.
We show that in the deconfined phase the axial symmetry remains unbroken to all 
orders in $1/N_c$.
In this case the breaking is due to instantons
which are described by spacelike D0-branes, in agreement with 't~Hooft's resolution.
The holographic dual of the symmetry breaking fermion condensate is a state
of spacelike strings between the D0-brane and the flavor D8-branes,
which result from a spacelike version of the string creation effect.
In the intermediate phase of deconfinement with broken chiral symmetry
the instanton gas approximation is possibly regulated in the IR, which would imply
an $\eta'$ mass-squared of order $\exp(-N_c)$.}

\begin{document}

\section{Introduction/history}

QCD has three light flavors of quarks and therefore an approximate global 
$SU(3)_R\times SU(3)_L$ chiral symmetry.
In nature this symmetry is spontaneously broken to the diagonal subgroup $SU(3)_V$,
and the corresponding pseudo-Goldstone bosons are identified with the eight light
pseudo-scalar mesons - the $\pi$'s, $K$'s and $\eta$.
The symmetry of the action is more precisely $U(3)_R\times U(3)_L$, and the unbroken
part is given by $SU(3)_V\times U(1)_V$, where $U(1)_V$ corresponds to Baryon number.
This would seem to imply another pseudo-Goldstone boson for the broken axial
symmetry $U(1)_A$, which is not observed. The candidate flavor-singlet $\eta'$ meson
is too massive. This is known as the $U(1)_A$ problem.

A resolution to this problem was first given by 't~Hooft \cite{'tHooft:1976up}.
The axial symmetry is broken by the anomaly, which is 
in turn related to instantons. Due to the presence of fermionic zero-modes in the instanton
background one needs to include fermions in the path integral in order to get a non-vanishing
amplitude. The correct combination is $\mbox{det}(\bar{\psi}_R \psi_L)$, where the determinant
is on the flavor indices, and the color indices are contracted in each bi-linear.
This is invariant under the chiral symmetry $SU(N_f)_R\times SU(N_f)_L$,
but not under the axial symmetry $U(1)_A$.
The instanton-induced condensate $\langle \mbox{det}(\bar{\psi}_R \psi_L) \rangle$ therefore
breaks $U(1)_A$ non-perturbatively. 
Summing over instantons and anti-instantons in a dilute
gas approximation then gives a new effective vertex in the action, which
leads to an $\eta'$ mass of order $\exp(-8\pi^2/g_{YM}^2)$.

However there are a couple of problems with the instanton picture.
The first is that the dilute instanton gas approximation breaks down when the
instantons become too large, and there is an IR divergence in the integral
over the instanton size. The second problem is that the instanton picture
appears to be in conflict with perturbative contributions to the $\eta'$ mass
in the quark model coming from quark-antiquark annihilation into gluons \cite{DeRujula:1975ge}.
This conflict was made sharper in the large $N_c$ limit by Witten \cite{Witten:1978bc}.
Perturbative effects lead to an $\eta'$ mass-squared
of order $1/N_c$ at large $N_c$, whereas the instanton gas picture would give a mass-squared
of order $\exp(-N_c)$. 
Witten argued that the conflict is resolved by confinement,
since in a confined gauge theory topological charge is not quantized, and it therefore
does not make sense to think about instantons, which are quanta of topological charge.
This lack of quantization is related to strong fluctuations of the topological charge density
in the confining vacuum, which is measured by the topological susceptibility of the
pure Yang-Mills theory $\chi_g= (d^2E/d\theta^2)_{\theta = 0}$. 
Witten and Veneziano derived the relation between this
quantity and the $\eta'$ mass \cite{Witten:1979vv,Veneziano:1979ec}
\be
\label{Witten-Veneziano}
m^2_{\eta'} = {2N_f\over f_\pi^2} \chi_g\,,
\ee
where $f_\pi$ is the pion decay constant.
Since $f_\pi^2\sim{\cal O}(N_c)$ one would get an order $1/N_c$ mass-squared if
$\chi_g\sim{\cal O}(1)$. Witten and Veneziano argued that although $\chi_g$ vanishes
order by order in perturbation theory, it could be finite in the large $N_c$ limit.



This proposal was confirmed using gauge/gravity duality.
In \cite{Witten:1998zw} Witten proposed a gravity dual for four-dimensional pure
Yang-Mills theory in terms of 4-branes in Type IIA string theory which are wrapped
on a circle with anti-periodic fermions. Using the duality he then showed that 
$\chi_g\sim {\cal O}(1)$ \cite{Witten:1998uk}.
More recently Sakai and Sugimoto proposed a model for massless QCD by
adding 8-branes and anti-8-branes to Witten's model \cite{Sakai:2004cn}.
The 8-branes and anti-8-branes
provide the right- and left-handed quarks. This model exhibits chiral symmetry 
breaking geometrically by the connection of the 8-branes to the anti-8-branes.
The mesons correspond to modes of the 8-branes in the connected configuration,
and one can compute their spectrum from the geometry.
In particular the $\eta'$ mass was shown to satisfy precisely (\ref{Witten-Veneziano})
with the $\chi_g$ computed in \cite{Witten:1998uk}.\footnote{For a description of 
$U(1)_A$-breaking in other gauge/gravity dual models see 
\cite{Armoni:2004dc,Barbon:2004dq}.}

Witten's model for pure Yang-Mills theory and Sakai and Sugimoto's extension
to massless QCD have also been analyzed at finite temperature \cite{Aharony:2006da}.
It was shown that the model undergoes deconfinement 
and chiral-symmetry-restoration transitions. 
One of the interesting features of this model is that 
for some range of the parameters chiral symmetry restoration occurs at a higher 
temperature than deconfinement (otherwise they happen at the same temperature).
In other words there is an intermediate phase where gluons are freed but quarks
remain bound in mesons.
In the deconfined phase one expects instantons to become relevant for
$U(1)_A$ breaking, and in particular
to contribute to the $\eta'$ mass in the intermediate phase.

The main goal of this paper is to understand whether and how 't~Hooft's resolution
is realized in the dual supergravity picture.
Using the supergravity description we will show that in the deconfined phase
$\chi_g$ vanishes to order 1 at large $N_c$, and $m^2_{\eta'}$ vanishes to order $1/N_c$.
Furthermore, since these results are based on a topological argument, we will argue
that they hold to all orders in the $1/N_c$ expansion.
The only contribution to $U(1)_A$ breaking in this phase must therefore come from instantons.
We will show that instanton charge, which corresponds in the supergravity picture to
to spacelike Euclidean 0-branes, is indeed quantized in this phase.
The presence of both the 0-brane and the 8-branes in the background leads to
a spacelike version of the familiar string-creation effect.
We will show that the resulting spacelike strings correspond precisely to the 
$U(1)_A$-breaking condensate $\langle \mbox{det}(\bar{\psi}_R \psi_L) \rangle$.
We will also argue that the IR problem in the instanton sum for computing the
mass of the $\eta'$ may be resolved by a cutoff at the chiral-symmetry breaking scale.
This implies an $\exp(-N_c)$
behavior for the $\eta'$ mass in the intermediate phase.




We begin in section 2 by briefly reviewing the models of Witten and Sakai-Sugimoto at
zero and finite temperature, with emphasis on the $U(1)_A$ issue.
In section 3 we show that $\chi_g=0$ to order 1 in the 
deconfined phase, and $m_{\eta'}=0$ to order $1/N_c$ in the 
intermediate phase of deconfinement with chiral symmetry breaking, and argue that
these hold to all orders in $1/N_c$.
In section 4 we discuss instantons in the dual supergravity picture, and show how 
they break $U(1)_A$ geometrically by a spacelike version of the string creation effect.
We have also included an appendix containing a worldsheet argument for the
spacelike string.



\section{Review of previous work}

\subsection{Holographic Yang-Mills}


The gravity dual of four-dimensional $SU(N_c)$ YM theory is constructed by wrapping
$N_c$ Type IIA D4-branes on a circle $x_4 \sim x_4 + \beta_4$ 
with anti-periodic boundary conditions
for the fermions, and taking the near-horizon limit \cite{Witten:1998zw}.
The background is given by \cite{Itzhaki:1998dd}
\be
\label{confined_background}
ds^2 &=& \left({u\over R}\right)^{3/2} \left(-dt^2 + \delta_{ij}dx^i dx^j
+ f(u)dx_4^2\right)
+ \left({u\over R}\right)^{-3/2}\left({du^2\over f(u)}
+ u^2 d\Omega_4^2\right) \nonumber \\[5pt]
e^{\Phi} &=& \left({u\over R}\right)^{3/4} \;,\;\;
F_4 = {3N_c\over 4\pi} \epsilon_4 \;,\;\;
f(u) = 1 - {u^3_{\Lambda}\over u^3} \,,
\ee
where $R=(\pi g_sN_c)^{1/3}l_s$, and $u_{\Lambda}$ is related to the periodicity
$\beta_4$ of $x_4$ as 
\be
 \beta_4 = {4\pi\over 3}\left({R^3\over u_{\Lambda}}\right)^{1/2} \,.
\ee
The four-dimensional YM coupling
is given by $g_{YM}^2=2(2\pi)^2 g_s l_s/\beta_4$ \cite{Aharony:1999ti},
and the 't~Hooft coupling is defined by $\lambda\equiv g_{YM}^2 N_c$.
The topology of the space is $\mathbb{R}^{3,1}\times D\times S^4$.
In particular the physical size of the circle goes to zero smoothly as $u$
approaches $u_{\Lambda}$ from above. 
This background gives confinement in
the dual gauge theory, which can be seen by computing the quark-antiquark potential,
corresponding to the action of a string with endpoints on the boundary at 
$u\rightarrow\infty$ \cite{Kinar:1998vq}. This can also be seen as a consequence
of the non-contractibility of the Polyakov loop in the Euclidean finite temperature
background. The role of the gauge theory parameter $\theta$ is played in the
dual background by the RR 1-form \cite{Witten:1998uk}
\be
\label{theta}
\theta = \int_{S^1} C_1\,,
\ee
where $S^1$ is the boundary at $u\rightarrow\infty$ of the disk $D$.
By Stokes' theorem it is therefore also related to the RR field strength $F_2=dC_1$ as
\be
\label{RR_flux}
\int_D F_2  =\theta \,.
\ee
A non-trivial vacuum angle $\theta$ therefore requires a non-trivial
RR field strength. The solution to the equation of motion for $C_1$ with
this condition for the flux on the disk is given by
\be
F_2 = {c\over u^4}\,\theta\, du\wedge dx_4 \,,
\ee
where $c=3u_\Lambda^3/\beta_4$.
Plugging this back into the kinetic term for $C_1$ in the supergravity action
and integrating over $u,x_4$ and the $S^4$
gives the four-dimensional energy density\footnote{The actual result is
$E(\theta)\propto\mbox{min}_k(\theta + 2\pi k)^2$, where $k$ is an integer.
This is because $\theta$ is an angle variable, whereas $F_2$ and $E$ are
real numbers.}
\be 
\label{vacuum_energy}
E(\theta) = {4\pi\over 3} {c\over (2\pi l_s)^6} \, \theta^2 = \chi_g \theta^2\,.
\ee
This confirms that $\chi_g\sim{\cal O}(1)$ at large $N_c$.

\subsection{The Sakai-Sugimoto model}

Massless quarks are included by adding $N_f$ 8-branes and $N_f$ anti-8-branes which are localized
in $x_4$ \cite{Sakai:2004cn}.\footnote{For other ways of adding flavor see
\cite{Karch:2002sh,Kruczenski:2003uq,Casero:2005se}.} 
For $N_f\ll N_c$ we regard the 8-branes as probes, and ignore their backreaction
on the background.\footnote{It's not
clear whether D8-branes can truly be treated in the probe approximation, since
their back-reaction on the background is significant. In particular the dilaton behaves
linearly in the coordinate transverse to the 8-branes, which makes it difficult to compactify
this coordinate. This is true even in the case with an equal number of 8-branes and
anti-8-branes. Although the RR tadpole cancels, the dilaton tadpole actually adds.}
This gives a $U(N_f)_R\times U(N_f)_L$ symmetry, which is global from the 4-brane point of view,
and new states transforming in the fundamental representations the gauge symmetry
and the global symmetry from the 4-8 and 4-$\bar{8}$ strings.
Since the 4-branes are extended along
$(x_1,\ldots,x_4)$ the 4-8 strings have $ND=6$ directions with mixed boundary
conditions. The NS sector of these strings is massive, and the R sector has a zero-energy
ground state with (complex) degeneracy $2^{4/2}=4$, corresponding to a massless 
Dirac fermion in the four-dimensional intersection of the branes.
The GSO projection leaves a Weyl fermion of one chirality.
For the 4-$\bar{8}$ strings the GSO projection is reversed, giving the
other chirality. The low energy states are therefore massless right-handed
fermions transforming as $({\bf{N_c}},{\bf{N_f}},{\bf 1})$ and massless left-handed
fermions transforming as $({\bf{N_c}},{\bf 1},{\bf{N_f}})$.
Since the 8-branes fill all the directions transverse to the 4-branes there is no obvious
way of giving an explicit mass to the fermions.

The most striking feature of this model is that it exhibits chiral symmetry breaking 
in a beautifully simple and intuitive way. 
In the decoupling limit the 4-branes are replaced with
their near-horizon background (\ref{confined_background}). Since the radial coordinate $u$ does
not extend down to the origin, the 8-branes and anti-8-branes must connect at some $u=u_0$
(fig.(1a)). The chiral symmetry of the 8-branes and anti-8-branes is therefore broken to the 
diagonal $U(N_f)_V$, and $u_0$ defines the scale of chiral symmetry breaking.
The connected configuration is a solution of the 8-brane DBI action in this background
corresponding to a U-shaped curve $\gamma(u,x_4)=0$ in the $(u,x_4)$ plane 
\cite{Sakai:2004cn,Aharony:2006da}. The minimal value of $u$ on the
8-branes $u_0$ is related to the asymptotic separation $L$ between the 8-branes and anti-8-branes.
At maximal (antipodal) separation $L=\beta_4/2$ the 8-branes connect at the minimal
value of the background $u_0=u_\Lambda$, and for small $L$ ($L\ll\beta_4$) $u_0\propto R^3/L^2$.

Mesons are described by modes of the 8-branes in the connected configuration.
In particular the pseudo-scalar mesons which correspond to the Goldstone bosons
of the broken chiral symmetry are identified with the 0-modes of the $u$ component of the 8-brane
worldvolume gauge field $A_u$ (we are using a parameterization in which
$u$ is a worldvolume coordinate). 
The $\eta'$ is in turn associated with $\mbox{tr}(A_u)$.
In the DBI action all of these modes, including $\eta'$, are massless.
However $\eta'$ becomes massive due to an anomaly in the bulk.
%
%
The gauge invariant RR field strength is shifted by the presence of the 8-branes to
\be
\label{shifted_RR}
\widetilde{F}_2 &=& dC_1 + i\, \mbox{tr}(A_u) \delta(\gamma(u,x_4)) \, du \wedge dx_4 \,.
\ee
Thus under a worldvolume gauge transformation the RR 1-form changes by
\be
\delta_\Lambda C_1 = -i\, \mbox{tr}(\Lambda) \delta(\gamma(u,x_4))\, dx_4 \,.
\ee
At $u\rightarrow\infty$ the 8-brane embedding curve $\gamma(u,x_4)$ reduces 
to $x_4=\mp L/2$, corresponding to the asymptotic positions of the 8-branes and anti-8-branes,
respectively.
It follows that $\theta$ transforms as
\be
\delta_\Lambda\theta = -i\, \mbox{tr}(\Lambda |_{x_4=-L/2} - \Lambda |_{x_4=L/2}) \,.
\ee
The RHS is the gauge transformation parameter in the relative overall $U(1)$ of the 
8-branes and anti-8-branes, which is precisely $U(1)_A$, and this 
is the correct transformation of $\theta$ under the axial symmetry.
Integrating (\ref{shifted_RR}) over the $(u,x_4)$ plane gives
\be
\label{shifted_RR_flux}
\int_D \widetilde{F}_2 = \theta + i\int_{-L/2}^{L/2} dx_4\, \mbox{tr} \left(
A_u \right) = \theta + {\sqrt{2N_f}\over f_\pi}\, \eta'\,,
\ee
where in the second equality we have used a relation between the $\eta'$ field
and the 8-brane gauge field which generalizes the one given in \cite{Sakai:2004cn}
for the special case of antipodal separation between the 8-branes and anti-8-branes
at $u\rightarrow\infty$. 
The supergravity action for the RR 1-form is modified,
\be
S_{C_1} = \int d^{10}x \sqrt{-g} |\widetilde{F}_2|^2 \,,
\ee
so the solution with the above condition on the flux is given by
\be
\label{shifted_RR_solution}
\widetilde{F}_2 = {c\over u^4}\left(\theta + \sqrt{2N_f\over f_\pi}\,\eta'\right) \, du \wedge dx_4 \,.
\ee
Plugging back into the action, and using the supergravity result for $\chi_g$
(\ref{vacuum_energy}), we see that $\eta'$ has a mass-squared
\be
m_{\eta'}^2 = {2N_f\over f_\pi^2} \chi_g \sim {\cal O}\left({1\over N_c}\right) \,,
\ee
in agreement with the Witten-Veneziano relation.

One comment we would like to add here is that although (\ref{shifted_RR_solution})
is a solution of the equation of motion $d*\widetilde{F}_2 = 0$, it does not in general
satisfy the modified Bianchi identity that follows from (\ref{shifted_RR}).
This is only satisfied for the constant mode of $\eta'$, which is however all we need to
compute the mass.

\begin{figure}
\centerline{\epsfig{file=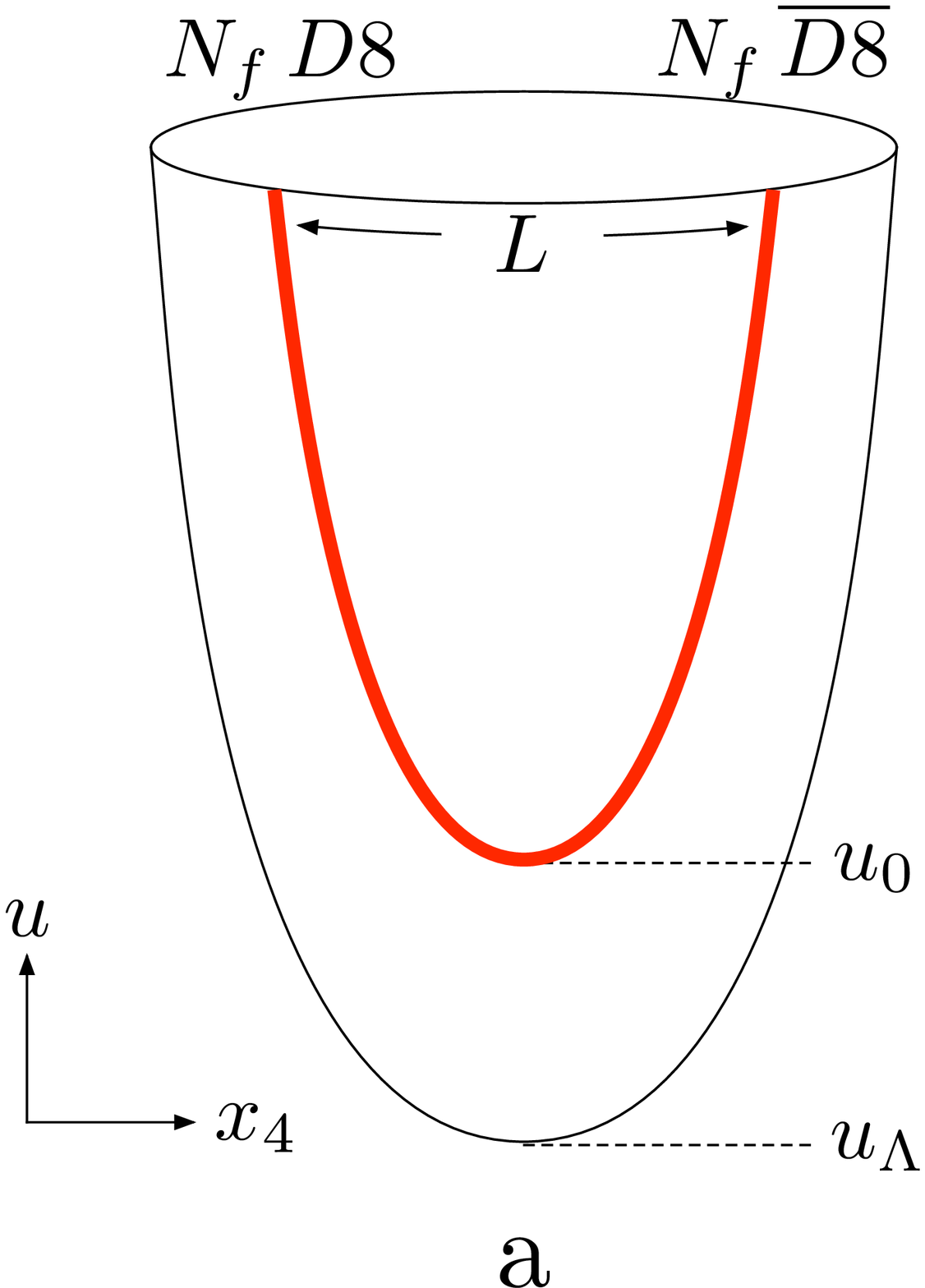,height=2in}\hspace{1cm}
\epsfig{file=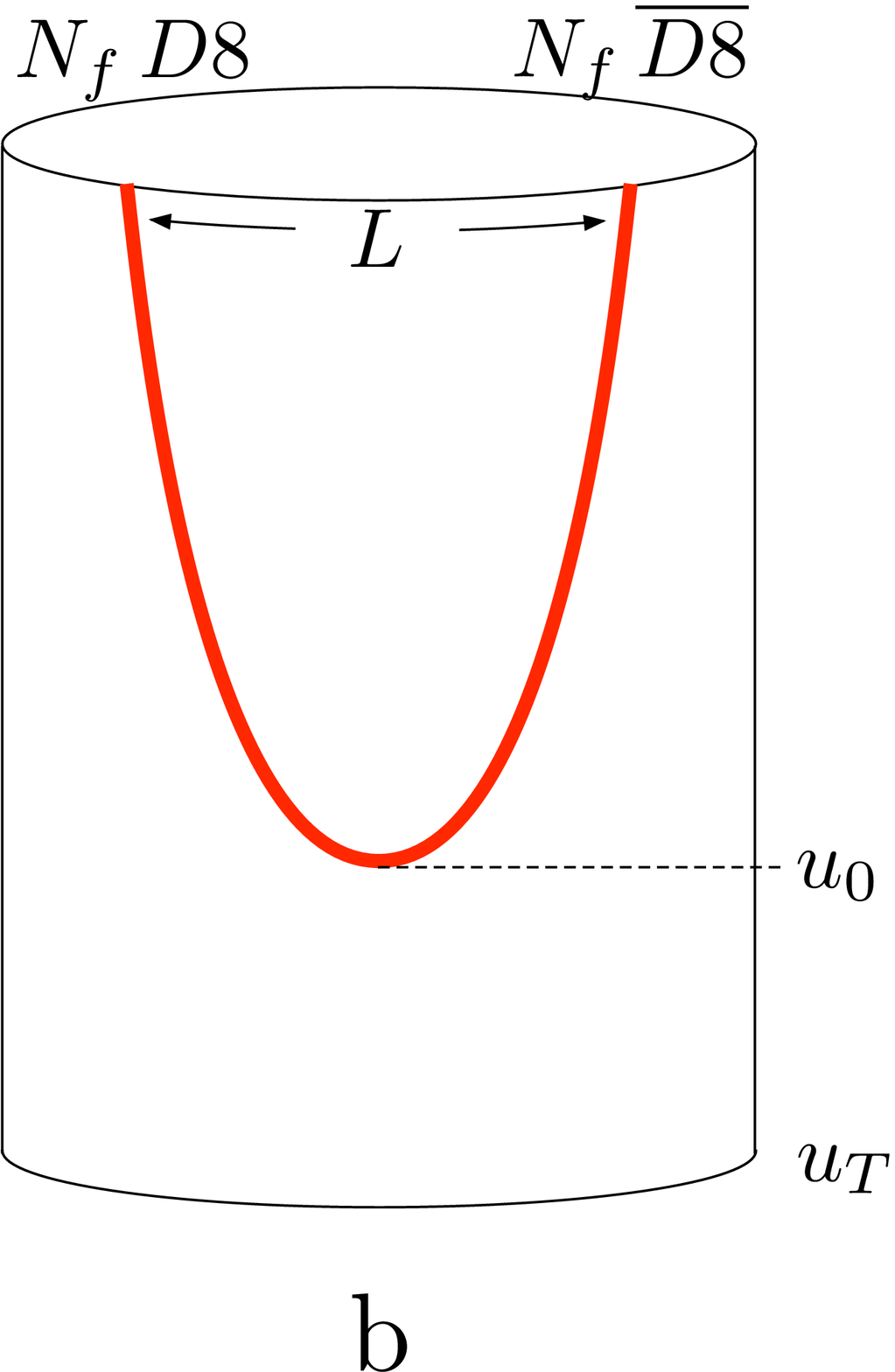,height=2in}\hspace{1cm}
\epsfig{file=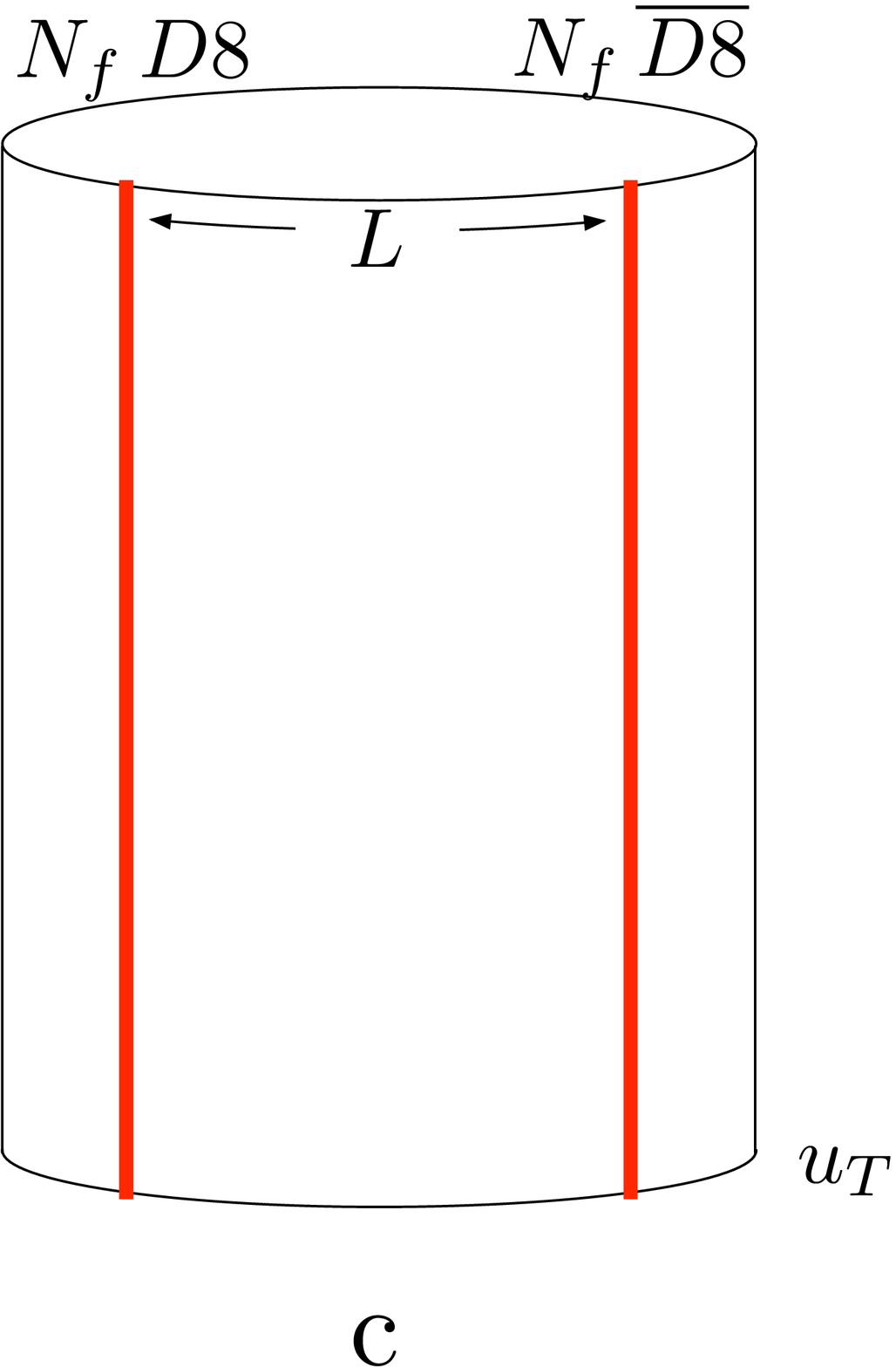,height=2in}}
\caption{The 8-brane-anti-8-brane configuration at finite temperature:
(a) confining phase (b) deconfinement with broken chiral symmetry 
(c) deconfinement with restored chiral symmetry.}
\end{figure}

\subsection{Finite temperature}

The finite temperature case was studied in \cite{Aharony:2006da}.
The finite temperature gauge theory is defined by replacing $t$ with $\tau=it$
and imposing the periodicity $\tau\sim\tau + \beta_\tau$, where $\beta_\tau = 1/T$.
That means that the boundary is topologically 
$\mathbb{R}^3\times S^1_{x_4}\times S^1_{\tau}\times S^4$.
There are two possible backgrounds with this boundary.
The first is just (\ref{confined_background}) with $t$ replaced by $\tau=it$.
The second is given by exchanging $\tau$ and $x_4$:
\be
\label{deconfined_background}
ds^2 = \left({u\over R}\right)^{3/2} \left(f(u)d\tau^2 + \delta_{ij}dx^i dx^j
+ dx_4^2\right)
+ \left({u\over R}\right)^{-3/2}\left({du^2\over f(u)}
+ u^2 d\Omega_4^2\right) .
\ee
In this background
\be
 f(u) = 1- {u_T^3\over u^3} \,,
\ee
where $u_T$ is related to the periodicity of $\tau$
\be
 \beta_{\tau} = {4\pi\over 3}\left({R^3\over u_{T}}\right)^{1/2} \,.
\ee
In this case the $x_4$ circle remains finite and the $\tau$ circle shrinks to zero size at $u=u_T$.
By comparing their free energies one can show that the first background is dominant at 
low temperatures, and the second background dominates at high temperatures.
The transition occurs at $T=1/\beta_4$.
In the gauge theory this is a confinement/deconfinement phase transition.
The background (\ref{confined_background}) describes the low-temperature
confined phase, and the background (\ref{deconfined_background}) corresponds
to the high-temperature deconfined phase.

With the flavor 8-branes and anti-8-branes
the model exhibits three phases in general. At low temperatures the 
background is (\ref{confined_background}) and the 8-branes and anti-8-branes
are necessarily in the connected configuration, so
both the gluons and the quarks are confined, and chiral symmetry is broken (fig.1a).
At high temperatures the background becomes
(\ref{deconfined_background}) and the gluons are deconfined. In this case
both the connected and disconnected 8-brane-anti-8-brane configurations are
solutions, and one must compare their free-energies to determine which is dominant.
At high enough temperatures the disconnected configuration wins and chiral symmetry
is restored (fig.1c). 
However for $L/R$ below a critical value ($\sim 0.97$) there is an intermediate phase
at $1/\beta_4 < T < 0.154/L$, in which the 8-branes and anti-8-branes remain
connected. In this phase chiral symmetry is broken even though the gluons are 
deconfined (fig.1b).


\section{Theta dependence and $\eta'$ mass in the deconfined phase}

Let us begin by repeating the computations of the topological susceptibility 
and the $\eta'$ mass for finite temperature.
Below the deconfinement transition both the background and the 8-brane configuration
are essentially unaltered from the zero temperature case. 
Therefore $\chi_g$ remains ${\cal O}(1)$ and
$m_{\eta'}^2$ remains ${\cal O}(1/N_c)$ in the confined phase.

In the deconfined phase the topology of the background changes:
the $(u,x_4)$ plane is a cylinder rather than a disk.
Therefore in the pure YM case we must replace 
the condition on the RR flux (\ref{RR_flux}) with
\be
\int_{C} F_2 = \theta - \int_{S^1_{u_T}} C_1 \,.
\ee
Now the minimal energy solution to the equation of motion is $F_2=0$, which
can be achieved by adjusting the second term on the right to equal $\theta$.
This shows that $\chi_g=0$ to order 1 in the deconfined phase.
Furthermore, since this argument relies only on the topology
of the background it is likely to hold to all orders in $1/N_c$.
This agrees with the expectation that in the absence of confinement the fluctuations of
the topological charge density should be suppressed, as can be seen for example in
lattice simulations \cite{Lucini:2004yh}.
Including the flavor 8-branes, then in the intermediate phase where the $\eta'$ exists
we must replace the condition on the shifted RR flux (\ref{shifted_RR_flux}) with
\be
\int_C \widetilde{F}_2 = \theta - \int_{S^1_{u_T}} C_1 + {\sqrt{2N_f}\over f_\pi}\, \eta' \,,
\ee
and therefore the minimal energy solution is $\widetilde{F}_2=0$.
(The same comment that was made at the end of section 2.2 applies here, namely that
this solution satisfies the modified Bianchi identity only for the constant mode of $\eta'$,
which is sufficient for computing the mass.)
We find that $m^2_{\eta'} = 0$, again to all orders in $1/N_c$.
As we will now see the $U(1)_A$ breaking and $\eta'$ mass in the deconfined phase
are due to instantons, precisely as originally proposed by 't~Hooft.


\section{Holographic description of $U(1)_A$ breaking}

\subsection{0-branes as instantons}

Yang-Mills instantons correspond to Euclidean D0-branes
whose worldlines wind around $x_4$  \cite{Witten:1998uk,Barbon:1999zp}.
These spacelike D0-branes couple to the $x_4$ component of $C_1$ 
which defines $\theta$.
For a single instanton, or a small number (relative to $N_c$) of instantons, we can treat
the D0-branes as probes in the D4-brane background.
The position of the D0-brane in ${\mathbb R}^4$ corresponds to the position
of the instanton, and the position of the D0-brane in the radial coordinate $u$
is related holographically to the size of the instanton $\rho$ 
as
\be
\rho^2 = {\beta_4 \lambda\over u} \,.
\ee

In the low-temperature background (\ref{confined_background}) the D0-brane is 
unstable (fig.2a) \cite{Barbon:1999zp}. It can unwind
since the circle is topologically trivial.
Furthermore, the D0-brane action in this background
is given by
\be
S_{D0} = m_{D0}\int_0^{\beta_4} dx_4 e^{-\Phi}\sqrt{g_{44}}
= {\beta_4\over g_s}\sqrt{1-{u_{\Lambda}^3\over u^3}} \,.
\ee
There is a $u$-dependent potential that pulls the D0-brane towards the tip
at $u_\Lambda$. The instanton therefore grows and disappears at $\rho\sim\beta_4$.
This is consistent with the large fluctuations of the topological charge density,
and confirms Witten's assertion that topological charge is not quantized in
the confining theory.

In the high temperature background dual to the
deconfined phase (\ref{deconfined_background})
the $x_4$ circle is topologically non-trivial, and the D0-brane is stable (fig.2b).
Its winding number corresponds to the instanton number.
This is consistent with the suppression of the topological charge density fluctuations
in this phase.
In this background the metric factor exactly cancels the dilaton factor and the action 
is independent of $u$,
\be
S_{D0} = {\beta_4\over g_s} = {8\pi^2 N_c\over\lambda}\,.
\ee
The instanton size $\rho$ is therefore a modulus in this phase.
The smallest radial position of the D0-brane is $u_T$, so 
instantons have a maximal size $\sim\beta_\tau$.


\begin{figure}
\centerline{\epsfig{file=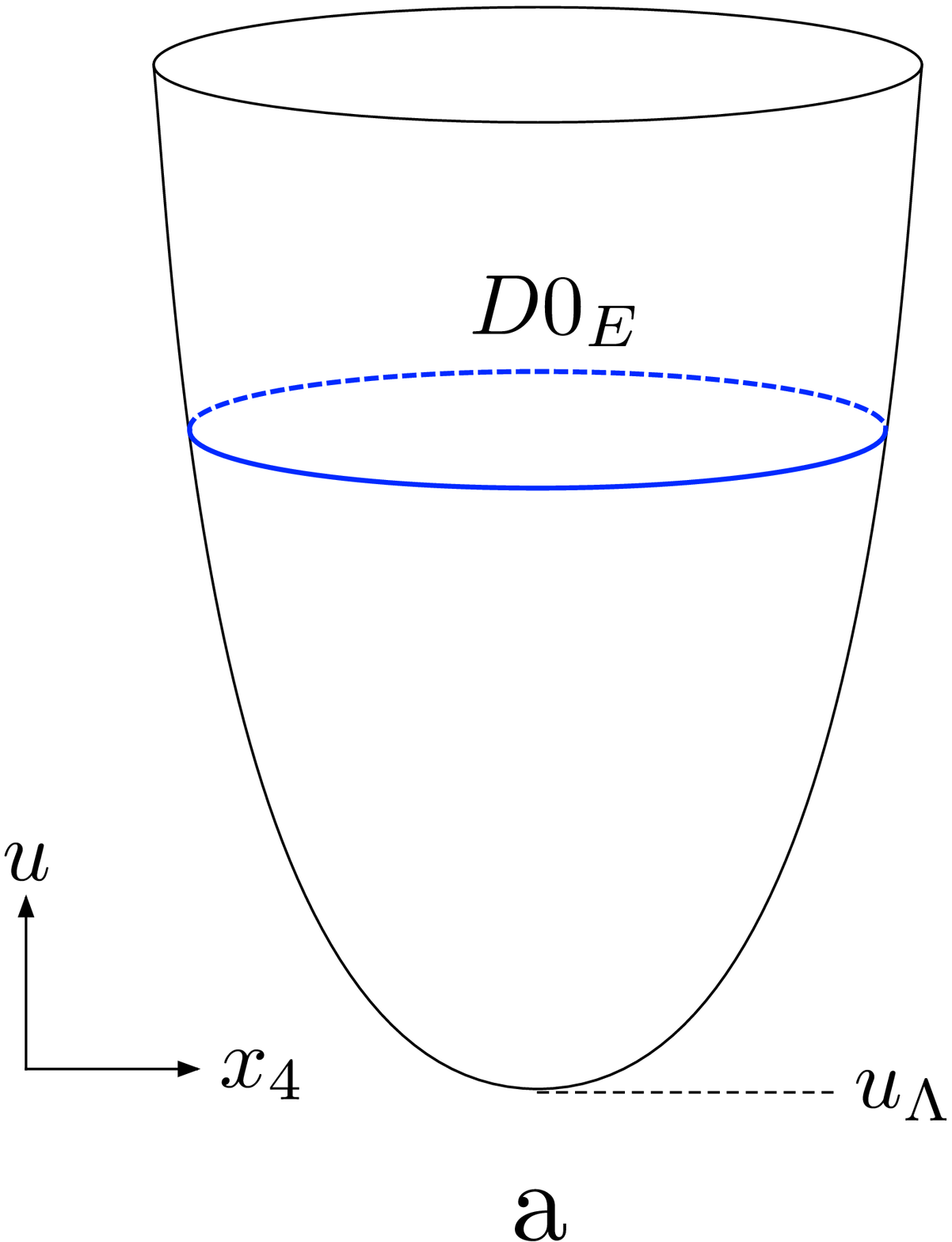,height=2in}\hspace{1cm}
\epsfig{file=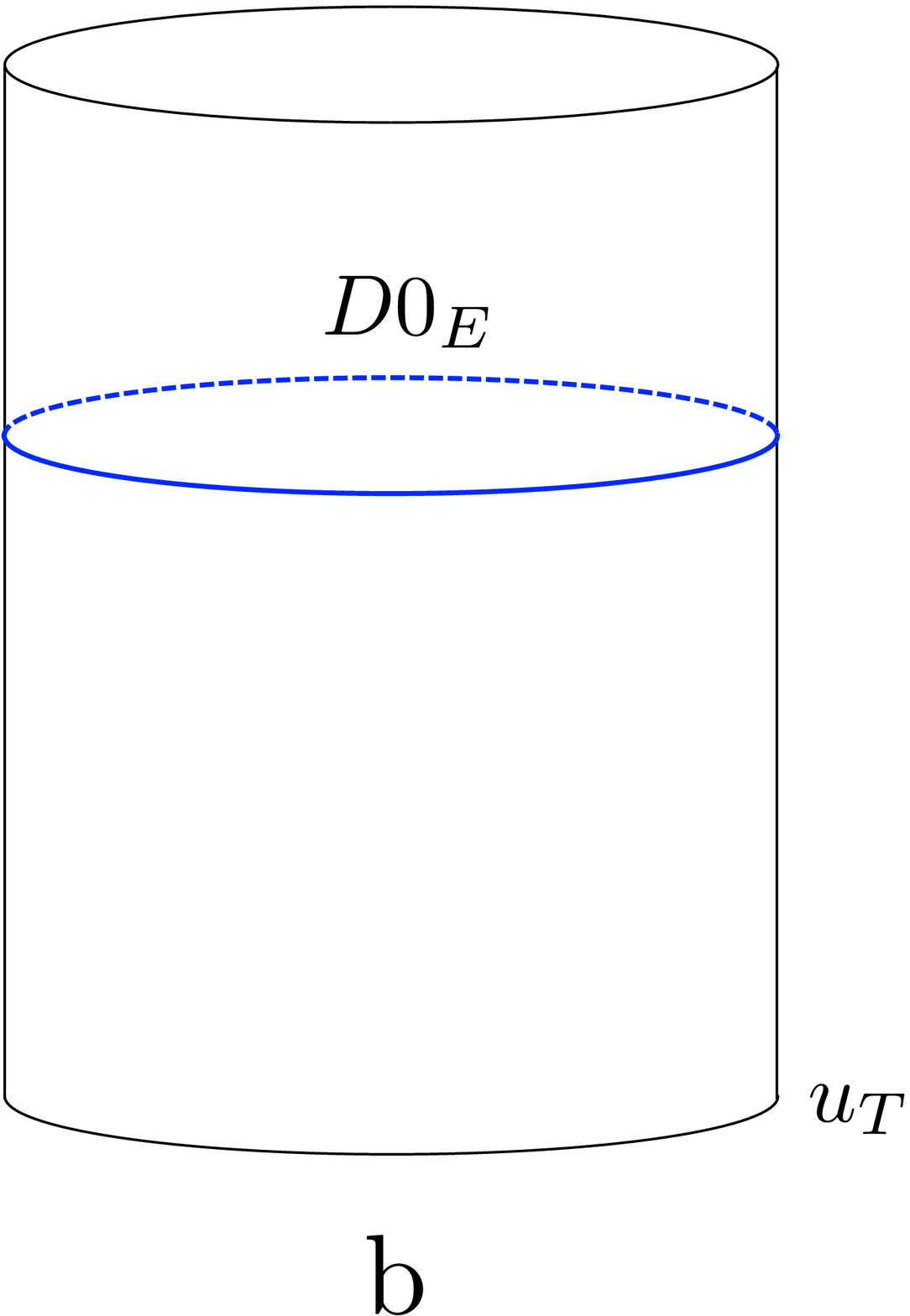,height=2in}}
\caption{Holographic instantons in YM theory: (a) confined phase
(b) deconfined phase}
\end{figure}

\subsection{Flavor branes and fermionic zero modes}

We now want to consider the effect of a YM instanton at large $N_c$ with $N_f$
flavors of massless quarks. In addition to the Euclidean D0-brane on $x^4$
we have the $N_f$ D8-branes and $N_f$ anti-D8-branes which are located at
$x^4=\mp L/2$, respectively. The corresponding instanton will have additional
degrees of freedom from the $0_E$-8 and $0_E$-$\bar{8}$ strings.
We will now show that these give the correct fermionic zero modes of
the instanton in the presence of the massless quarks.

The $0_E$-8 string is an example of an $ND=10$ string, {\em i.e.} it has mixed 
boundary conditions in all directions. The bosonic oscillators are all half-odd-integer moded,
and the fermionic oscillators are all integer moded in the NS sector, and half-odd-integer
moded in the R sector. Thus the roles of the NS and R sector fermions are reversed relative
to the $ND=0$ case of identical parallel D-branes. In particular, before the GSO projection, 
the spacetime-bosonic NS ground state is 32-fold degenerate, and the spacetime-fermionic
R ground state is non-degenerate. The former is massive, and the latter, as usual, is massless.
The action of $(-1)^F$ is the same as in the $ND=0$ case, but with R and NS
reversed:
\be
(-1)^F |0\rangle_R &=& - |0\rangle_R \nonumber\\
(-1)^F |{\bf s}\rangle_{NS} &=& |{\bf s'}\rangle_{NS} \Gamma_{\bf s's} \,.
\ee
The GSO projection ${1\over 2}(1+(-1)^F)$ therefore removes the massless fermion from
the spectrum of the $0_E$-8 string 
(just like it removed the tachyon from the spectrum of the $ND=0$ string).
On the other hand if we replace one of the two branes with an antibrane this reverses
the GSO projection to ${1\over 2}(1-(-1)^F)$, and the massless fermion remains
(like the tachyon in the brane-antibrane system).
So the massless spectrum of this string includes a single complex fermion
degree of freedom for $0_E$-$\bar{8}$ and 
$\bar{0}_E$-$8$,
and none for $0_E$-$8$ and 
$\bar{0}_E$-$\bar{8}$. In the D4-brane background
this gives the correct fermionic zero modes for the instanton and anti-instanton (according to
the Atiyah-Singer index theorem).
In the instanton case the zero mode is associated with the anti-8-brane, and therefore
corresponds to an on-shell left-handed fermion in four dimensions.
In the anti-instanton case the zero mode comes from the 8-brane, and therefore
corresponds to an on-shell right-handed fermion in four dimensions.
For $N_f$ 8-branes we get the required multiplicity of $N_f$ fermionic zero modes.
Note that in the supergravity picture the fermionic zero modes are separate from the
quarks. The latter come from the 4-8 and 4-$\bar{8}$ strings and the former from
the $0_E$-$\bar{8}$ or $\bar{0}_E$-8 strings.

The presence of the fermionic zero modes in the gauge theory implies a 
non-trivial multi-fermion condensate which breaks the axial symmetry.
How can we interpret this condensate in the dual supergravity picture?
As we will now explain the condensate $\langle \mbox{det}(\bar{\psi}_R \psi_L) \rangle$
has a very simple geometrical description in the dual background: it is a state
of spacelike strings which are enclosed by the 0-brane, 8-branes
and anti-8-branes.

\subsection{Spacelike strings}

Consider first the $0_E$-8 configuration in flat space.
As it stands this configuration is incomplete: it requires a semi-infinite
spacelike string worldsheet ending on the 0-brane worldline on one side
of the 8-brane. This can roughly be thought of as a "superluminal" version of the 0-8
string creation effect 
\cite{Hanany:1996ie,Bachas:1997ui,Danielsson:1997wq,Bergman:1997gf}. 
It can be proven by an argument similar to the one
given in \cite{Polchinski:1995sm} for the ordinary 0-8 configuration. 
The 8-brane is a source for the
RR 0-form field strength $F_0$ (the Poincare dual of the 10-form $F_{10}$),
such that its value jumps by 1 across the 8-brane. We adopt the convention that
$F_0=0$ to the left of the 8-brane and $F_0=1$ to the right (this can be achieved
by placing another 8-brane at infinity on the left). For the anti-8-brane the left
and right sides are reversed. 
A non-zero $F_0$ is necessarily a background of massive Type IIA supergravity
\cite{Romans:1985tz}, in which the (sourceless) equation of motion for the $B$ field is
\be
d(*H) = F_0 *F_2 \,.
\ee
Now consider a spacelike 0-brane along $x_4$ in this background, and integrate the above equation 
over an $S^8$ surrounding the 0-brane in Euclidean spacetime. 
The LHS vanishes, whereas the RHS gives $F_0$.
We therefore need to include a source term of strength $F_0$ on the RHS, such that
its integral on the $S^8$ cancels this. Geometrically this is a spacelike string worldsheet
which is wedged between the 0-brane and the 8-brane on the $F_0=1$ side (fig.3a).
For the anti-8-brane the worldsheet is on the other side (fig.3b), and therefore 
in an 8-$\bar{8}$ configuration the worldsheet is enclosed between
them (fig.3c). In the appendix we give an alternative worldsheet argument for 
the existence of this string.

\begin{figure}
\centerline{\epsfig{file=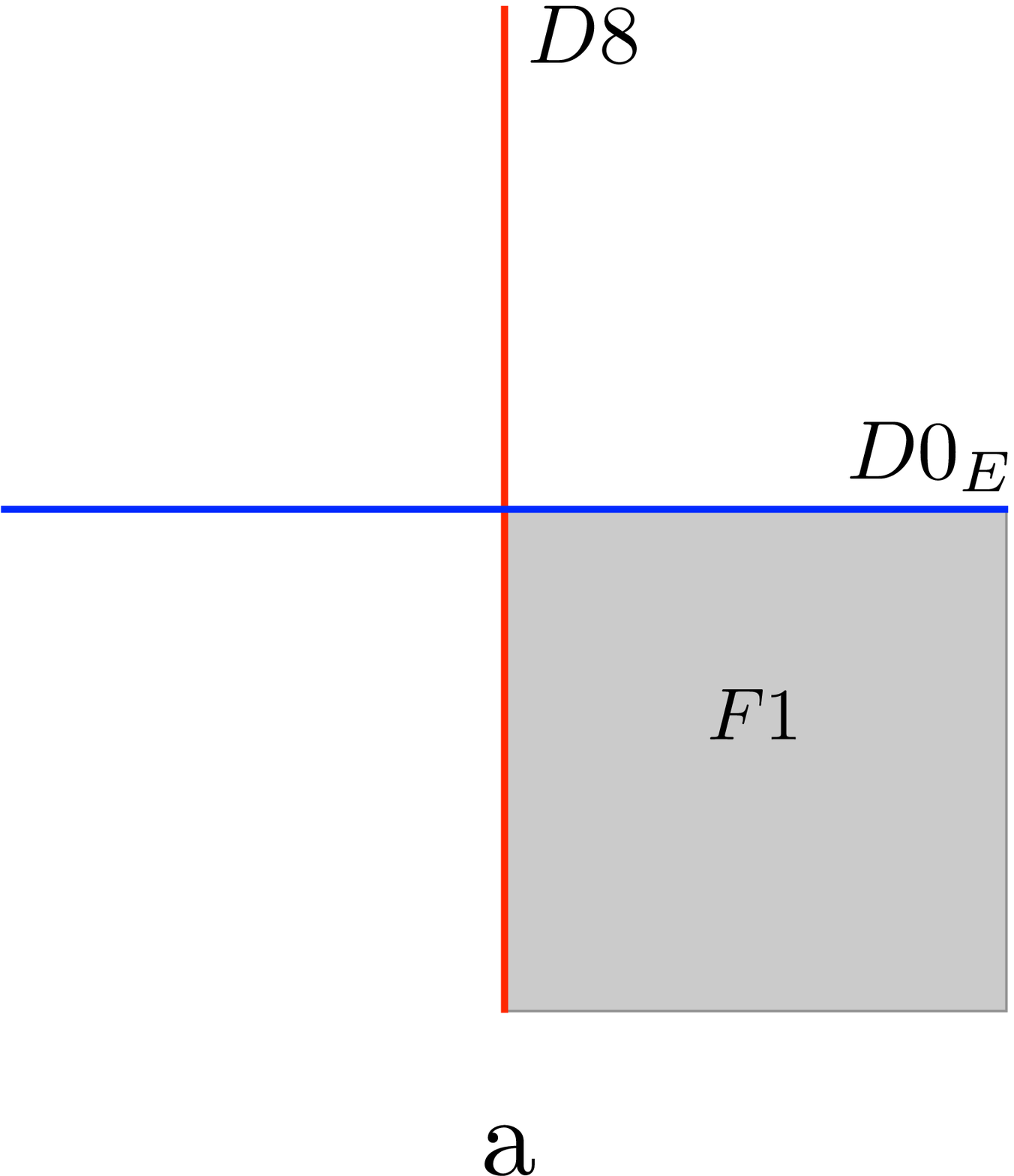,width=3.5cm}\hspace{2cm}
\epsfig{file=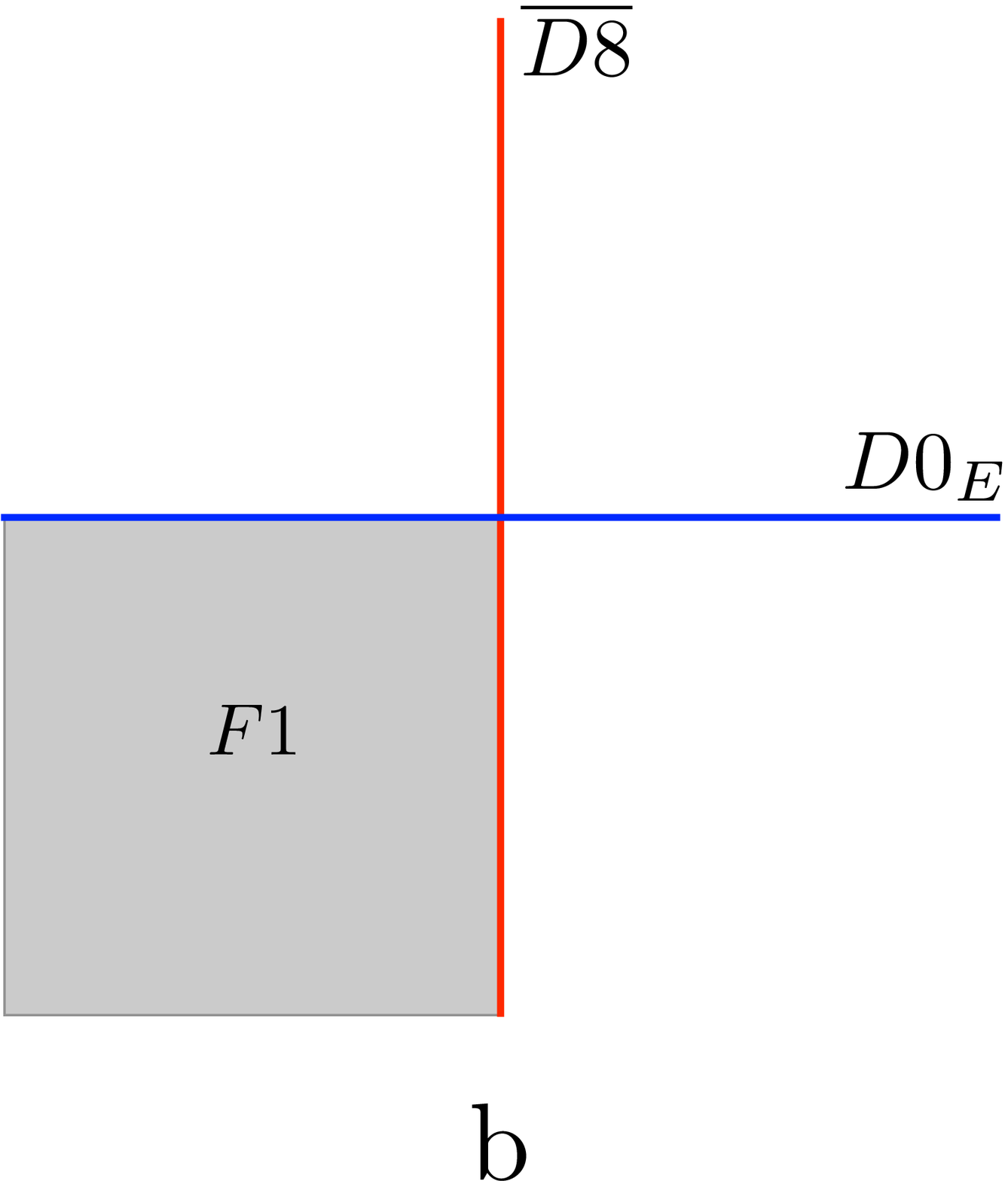,width=3.5cm}\hspace{2cm}
\epsfig{file=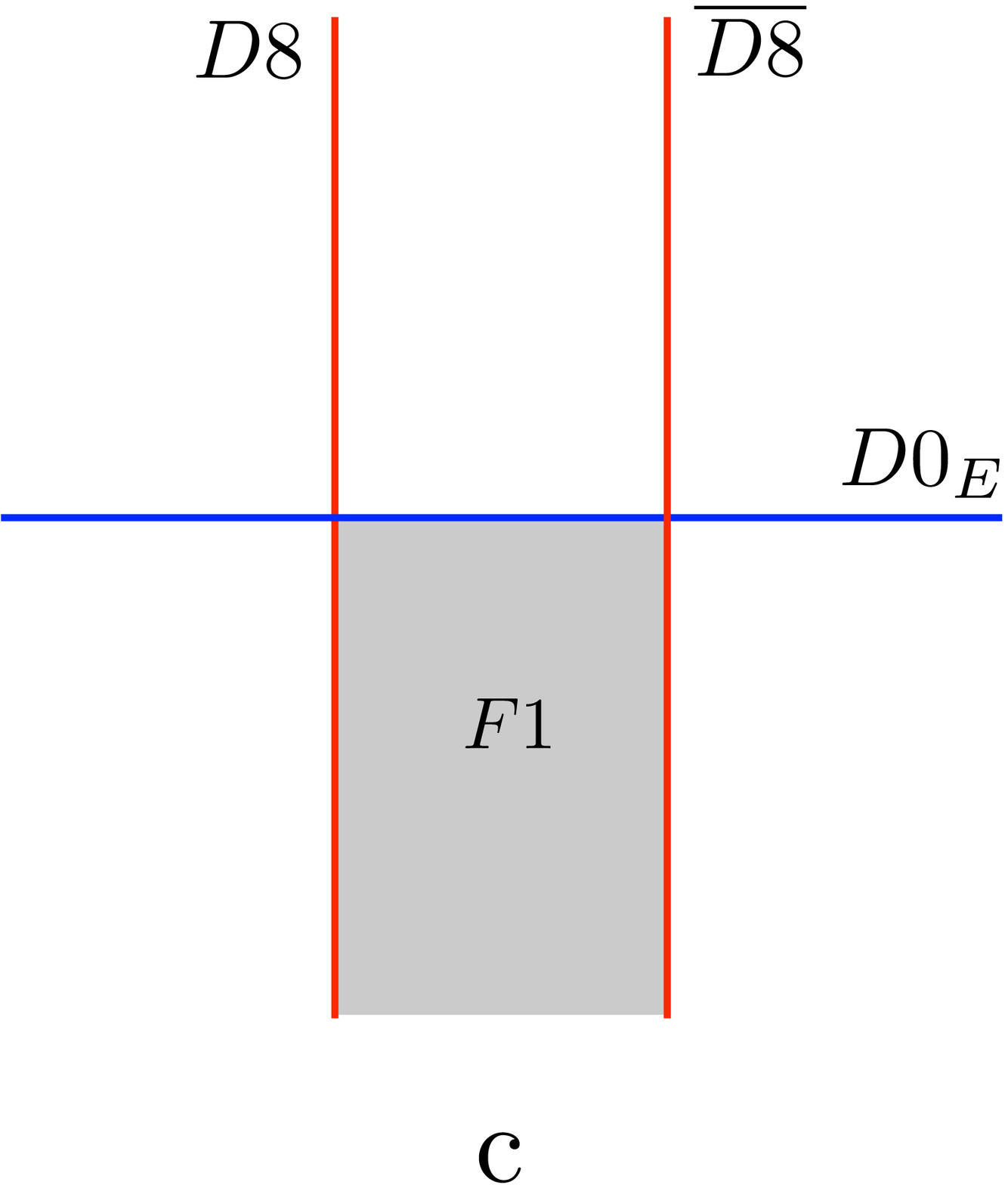,width=3.5cm}}
\caption{The correct D$0_E$-D8-string configurations.}
\end{figure}

\subsection{Breaking $U(1)_A$ holographically}

Applying this reasoning to the situation at hand we see that the configuration
must include a spacelike string for each 8-brane-anti-8-brane pair, which,
for the lowest energy,
fills the region in the $(u,x_4)$ plane bounded by the 0-brane and the
8-brane-anti-8-brane pair as shown in figure 4.
We will now argue that these $N_f$ spacelike strings are the holographic dual 
of the axial symmetry breaking condensate $\langle \mbox{det}(\bar{\psi}_R\psi_L)\rangle$.
To this end we will determine the $U(N_f)_R\times U(N_f)_L$ quantum numbers of the 
spacelike string state, and show that they agree with those of the condensate.

The enclosed spacelike string can be quantized using two different parameterizations.
Consider first the parameterization where one endpoint of the string is on the 0-brane
boundary, and the other endpoint is on the anti-8-brane boundary (in the connected configuration
this means on the right half of the 8-brane curve). In this parametrization the worldsheet
is foliated by $0_E-\bar{8}$ strings (fig.5a). The ground state of this string is a single massless
fermion. Therefore for $N_f$ $\bar{8}$-branes (and $N_f$ 8-branes) the state must be
anti-symmetrized in the anti-8-brane Chan-Paton index. 
Now consider a different parameterization
of the worldsheet where one endpoint is on the 8-brane boundary
and the other endpoint is on the anti-8-brane boundary (fig.5b).
In this case the ground state of the open string is the tachyon.
For $N_f$ 8-branes and anti-8-branes there are $N_f^2$ tachyons $T^{ij}$.
As we saw in the first parameterization the state must be anti-symmetric in one
of the indices, in other words it has the form 
$\epsilon_{j_1\cdots j_{N_f}} T^{i_1j_1}\cdots T^{i_{N_f}j_{N_f}}$. 
Since $T^{ij}$ is bosonic the other index is also anti-symmetrized and we get
$\epsilon_{i_1\cdots i_{N_f}} \epsilon_{j_1\cdots j_{N_f}} T^{i_1j_1}\cdots T^{i_{N_f}j_{N_f}}
 = \mbox{det}\, T$. This has precisely the quantum numbers of the fermionic condensate,
 so we identify it as the holographic dual,
 \be
 \mbox{det}\, T \leftrightarrow \langle \mbox{det}(\bar{\psi}_R\psi_L)\rangle \,.
 \ee
This identification is also consistent with the identification of the tachyon $T^{ij}$ 
as the dual the fermion bilinear $\langle\bar{\psi}_R^i\psi_L^j\rangle$, which is responsible
for breaking the chiral symmetry \cite{Antonyan:2006vw}.
We conclude that the anti-symmetrized state of spacelike strings is what is
responsible for breaking $U(1)_A$ in the dual supergravity picture.

\begin{figure}
\centerline{\epsfig{file=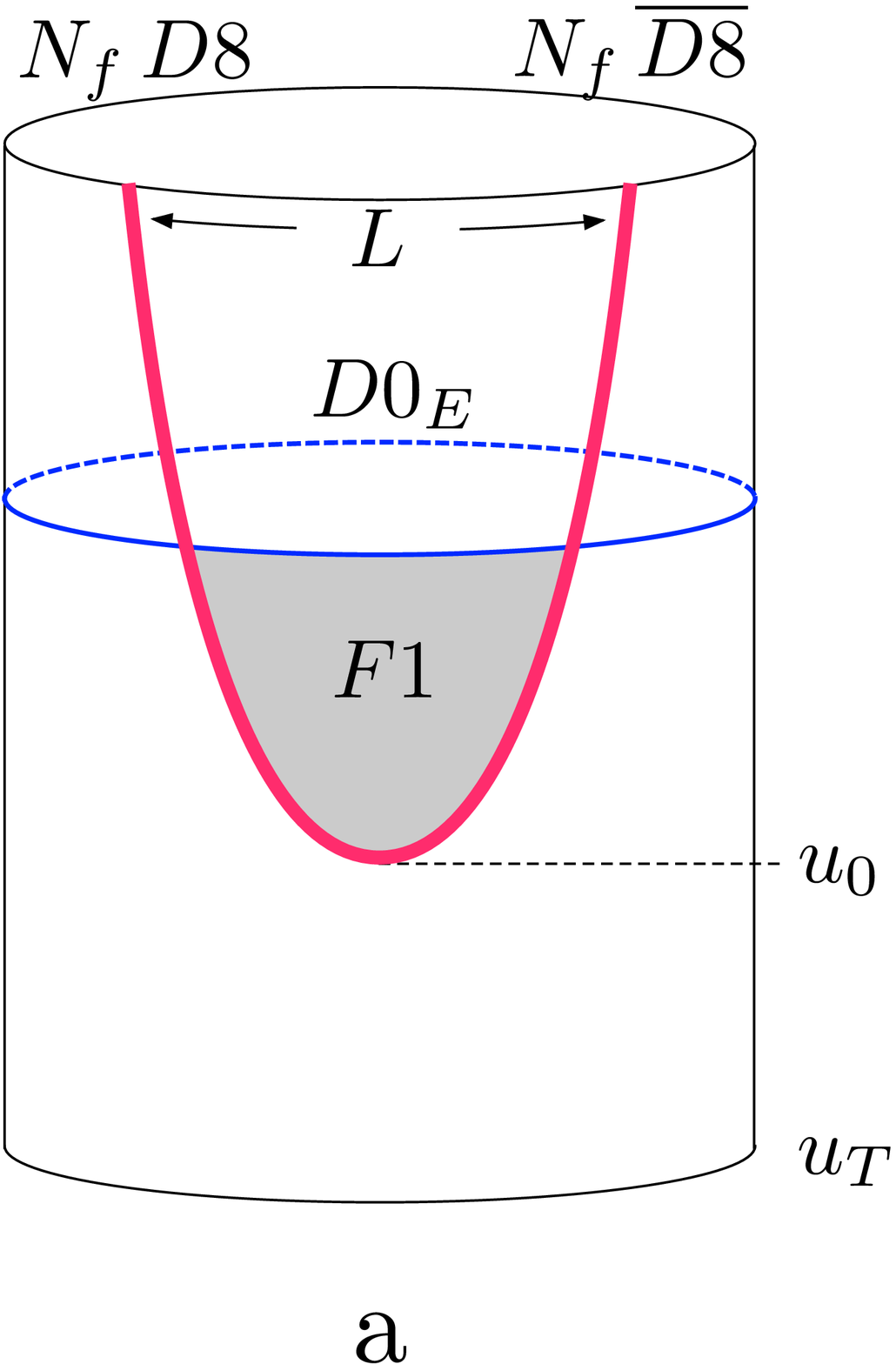,height=2in}\hspace{1cm}
\epsfig{file=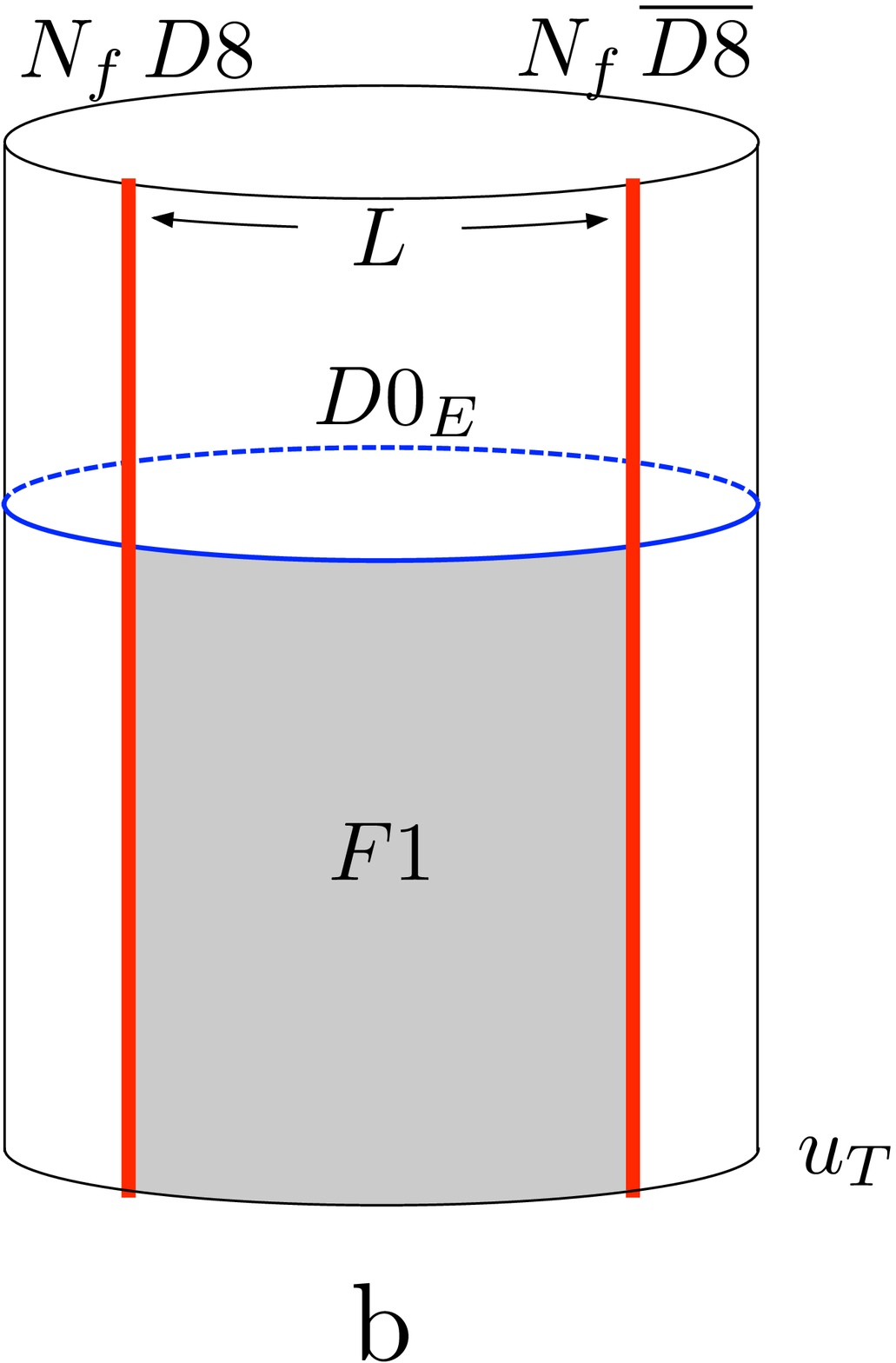,height=2in}}
\caption{Instanton and spacelike strings: (a) broken chiral symmetry phase,
(b) restored chiral symmetry phase.}
\end{figure}

\begin{figure}
\centerline{\epsfig{file=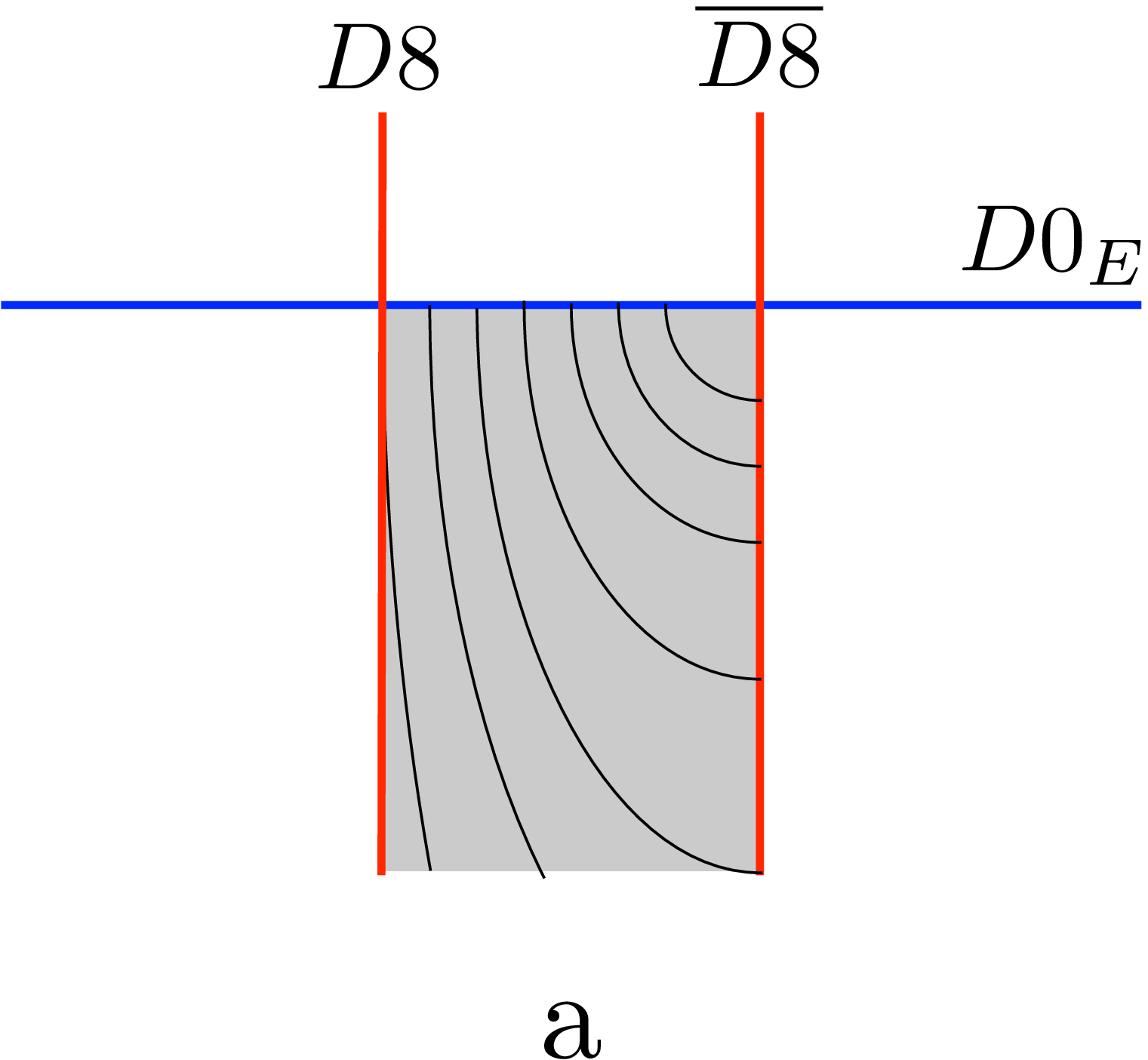,height=3cm}\hspace{1cm}
\epsfig{file=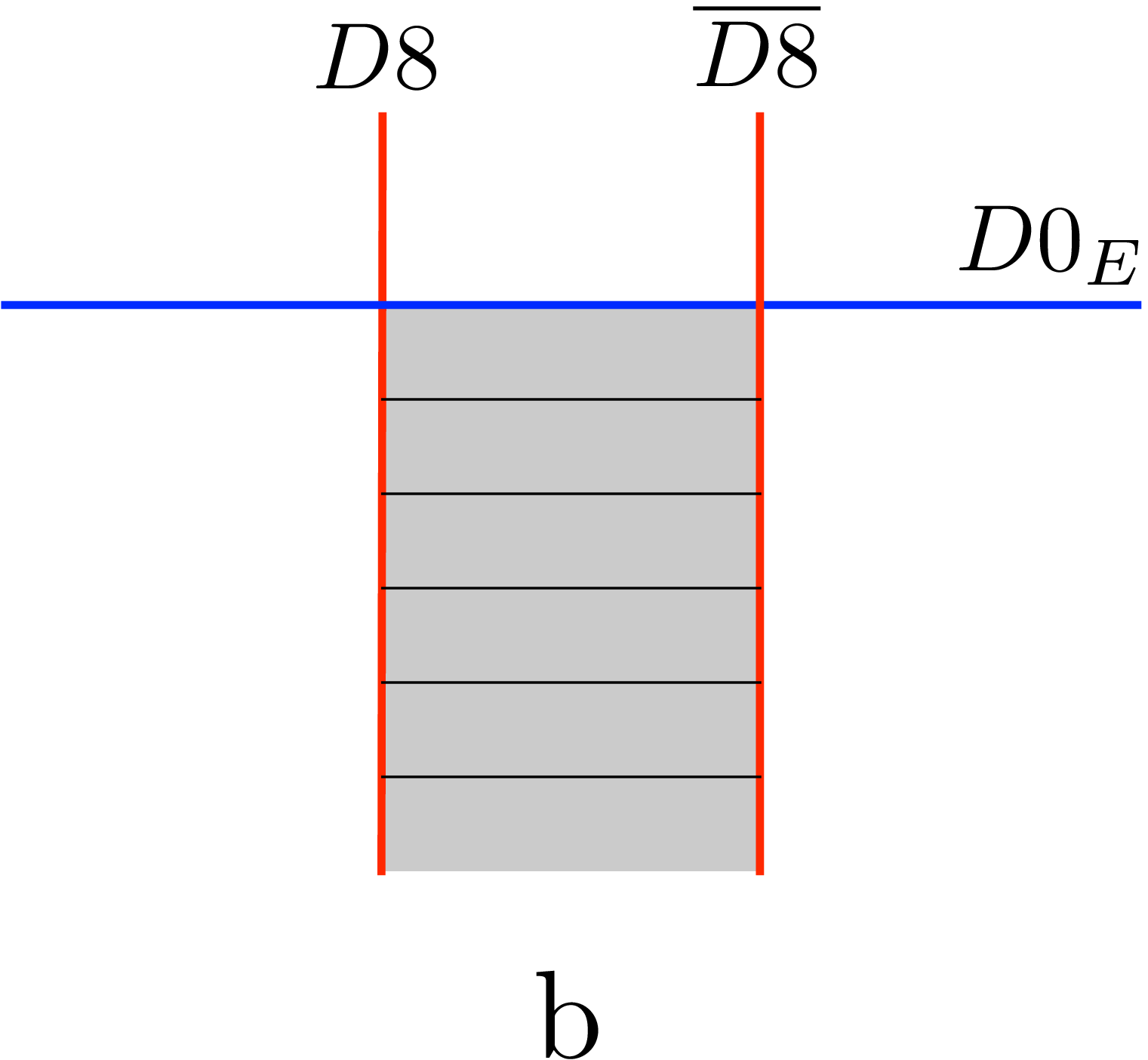,height=3cm}}
\caption{Parameterizations of the spacelike string:
(a) foliating with D$0_E$-$\overline{\mbox{D}8}$ strings,
(b) foliating with D8-$\overline{\mbox{D}8}$ strings.}
\end{figure}

\subsection{IR regulated instanton gas?}

In computing the instanton contribution to the fermionic condensate one has to integrate
over the instanton moduli space, including the instanton size $\rho$. This integral is IR
divergent. Furthermore, in summing over the contributions of multiple instantons and
anti-instantons one assumes a dilute gas approximation, which breaks down
when the instantons become too large. Finite temperature provides a natural IR cutoff
on the instaton size integral. We can see this explicitly in the dual background fig.2b,
where there is a minimal value of $u$ for the 0-brane position. However this will not by itself
solve the second problem, since a gas of instantons of size on the order of the cutoff will
not be dilute. A possible solution to this problem can be seen in the intermediate phase
(fig.4a), which is of-course the one relevant for the $\eta'$ mass computation.


In this phase the 8-branes and anti-8-branes
connect at $u=u_0\geq u_T$ and the 0-brane crosses them only for $u>u_0$. 
For 0-branes below $u_0$ the fermionic zero modes are lifted and there is no
spacelike string. 
This is qualitatively what one expects in the gauge theory, since $u_0$ is the energy scale of chiral symmetry
breaking. Above this scale there are free massless quarks and chiral symmetry is restored, so
the instanton should have $N_f$ fermionic zero modes. Below this scale chiral symmetry
is broken and quarks and anti-quarks bind into mesons, so the fermionic zero modes
should be lifted. 
Therefore instantons larger than $\rho_c=\sqrt{\beta_4\lambda/u_0}$ do not
contribute to $U(1)_A$ breaking. Since this is smaller than the maximal size 
$\rho_m=\sqrt{\beta_4\lambda/u_T}$ the dilute gas approximation might actually be
valid in this phase. If so, it would imply an $\eta'$ mass $m^2_{\eta'}\sim{\cal O}(e^{-N_c})$
in the intermediate phase.


\section*{Acknowledgments}
We would like to thank Cobi Sonnenschein and Shigeki Sugimoto for 
helpful conversations.
This work was supported in part by the
Israel Science Foundation under grant no.~568/05.
O.B. would also like to thank the Aspen Center for Physics where
part of this work was done.

\appendix

\section{A worldsheet argument for the spacelike string}

Another way to deduce the existence of the spacelike string in the
$0_E$-8 configuration is by computing the vacuum annulus diagram
of the $0_E$-8 string. The analogous computation in the 0-8 case was 
done in \cite{Lifschytz:1996iq,Bergman:1997gf}.
In the present case the string has $ND=10$, which means all the bosonic 
oscillators are half-odd-integer moded, and the fermionic oscillators are all
integer moded in the NS sector and all half-odd-integer moded in the R sector.
The R ground state is therefore non-degenerate (and massless).
The ghosts and superghosts have the usual moding.
Defining as usual $q=e^{-\pi t}$ and 
\begin{eqnarray}
f_{1}(q)&=&q^{1/12}\Pi_{n=1}(1-q^{2n})\\
f_{2}(q)&=&\sqrt{2}q^{1/12}\Pi_{n=1}(1+q^{2n})\\
f_{3}(q)&=&q^{-1/24}\Pi_{n=1}(1+q^{2n-1})\\
f_{4}(q)&=&q^{-1/24}\Pi_{n=1}(1-q^{2n-1}) \,,
\end{eqnarray}
the vacuum annulus amplitude can be written as
\begin{equation}
\label{annulus}
A_{0_E-8}=\int_{0}^{\infty}\frac{dt}{t}\frac{f_{1}^{2}}{f_{4}^{10}}
\left[-\frac{f_{3}^{10}}
{f_{2}^{2}}+\frac{f_{2}^{10}}{f_{3}^{2}}\pm\frac{f_{4}^{10}}{f_{1}^{2}}\cdot\infty
\right] \,.
\end{equation}
The sign in the third term is associated with the sign in the GSO projection
(in the R sector),
and therefore depends on whether we have an 8-brane or an anti-8-brane,
and the infinite factor comes from the trace over the superghost zero modes. 
We can now think of a process where we rotate the 8-brane such that it
crosses the $0_E$-brane. Once rotated by $\pi$ the 8-brane becomes
an anti-8-brane, so the sign in (\ref{annulus}) flips. The difference between
the initial and final amplitudes is infinite, and is accounted for by the creation
of a spacelike string (fig.6).
In fact this is just the ordinary string creation effect: when the 8-brane crosses
the $0_E$-brane they are parallel, and the configuration is equivalent to the
0-8 system. 
The divergence in this case is associated with the infinite
extent of the spacelike worldsheet of the string.
To make the picture symmetric between the 8-brane and anti-8-brane cases we need 
to start with a worldsheet in the lower-right quadrant for the 8-brane, which has 
the opposite orientation of the one which is created in the crossing.
The configuration after the rotation will have an anti-8-brane with a worldsheet in the
lower-left quadrant (fig.3).

\begin{figure}
\centerline{\epsfig{file=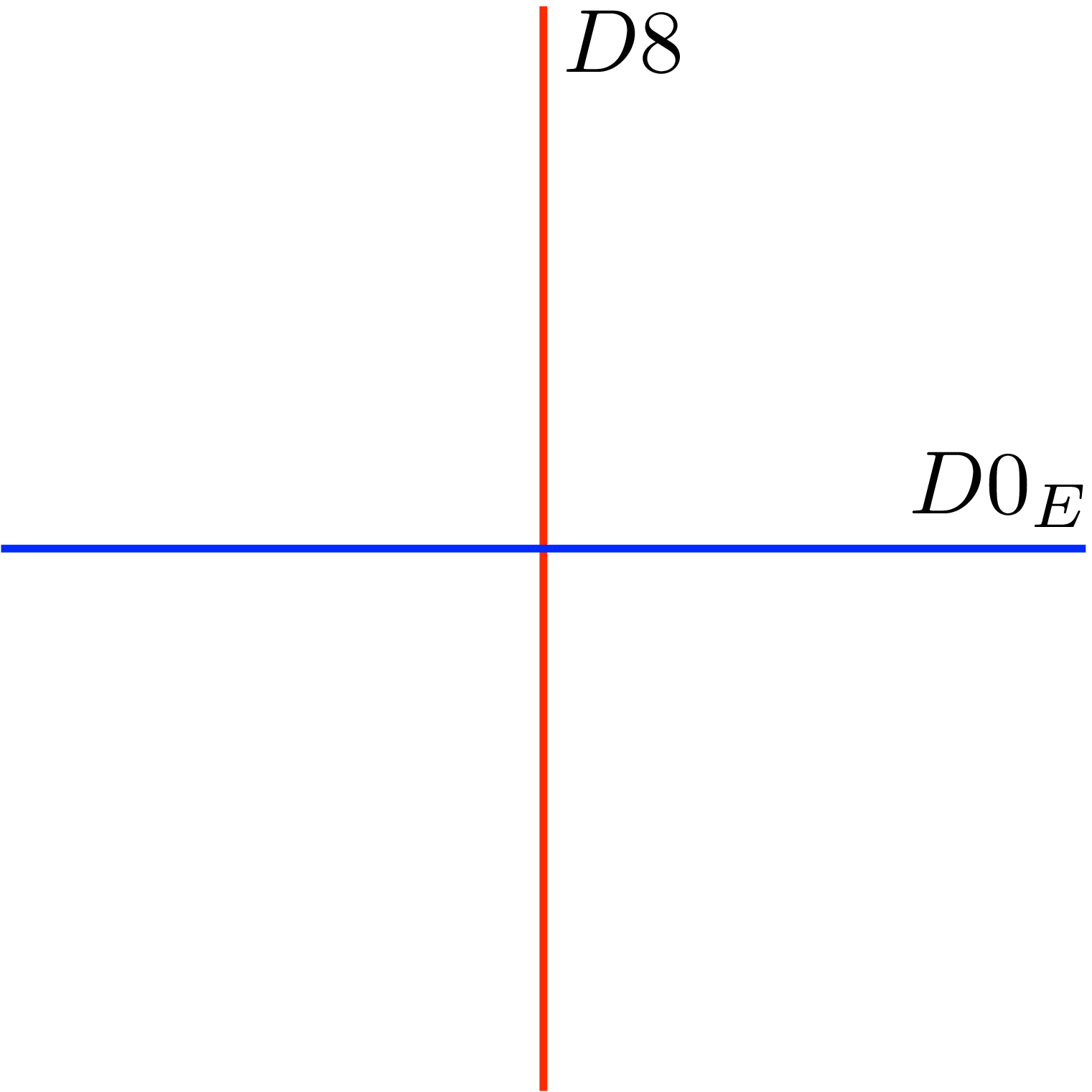,width=3.5cm}\hspace{0.5cm}
\epsfig{file=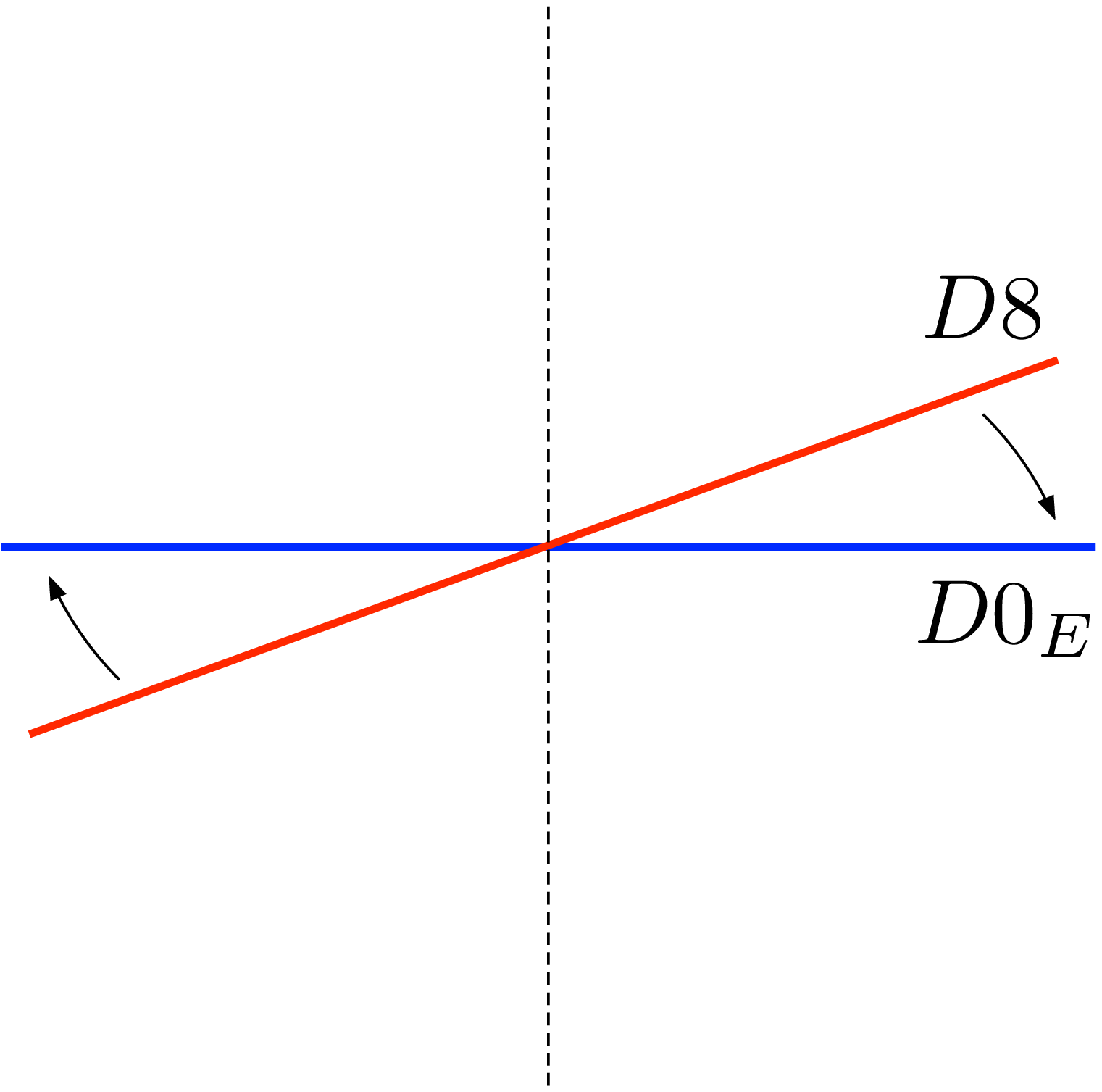,width=3.5cm}\hspace{0.5cm}
\epsfig{file=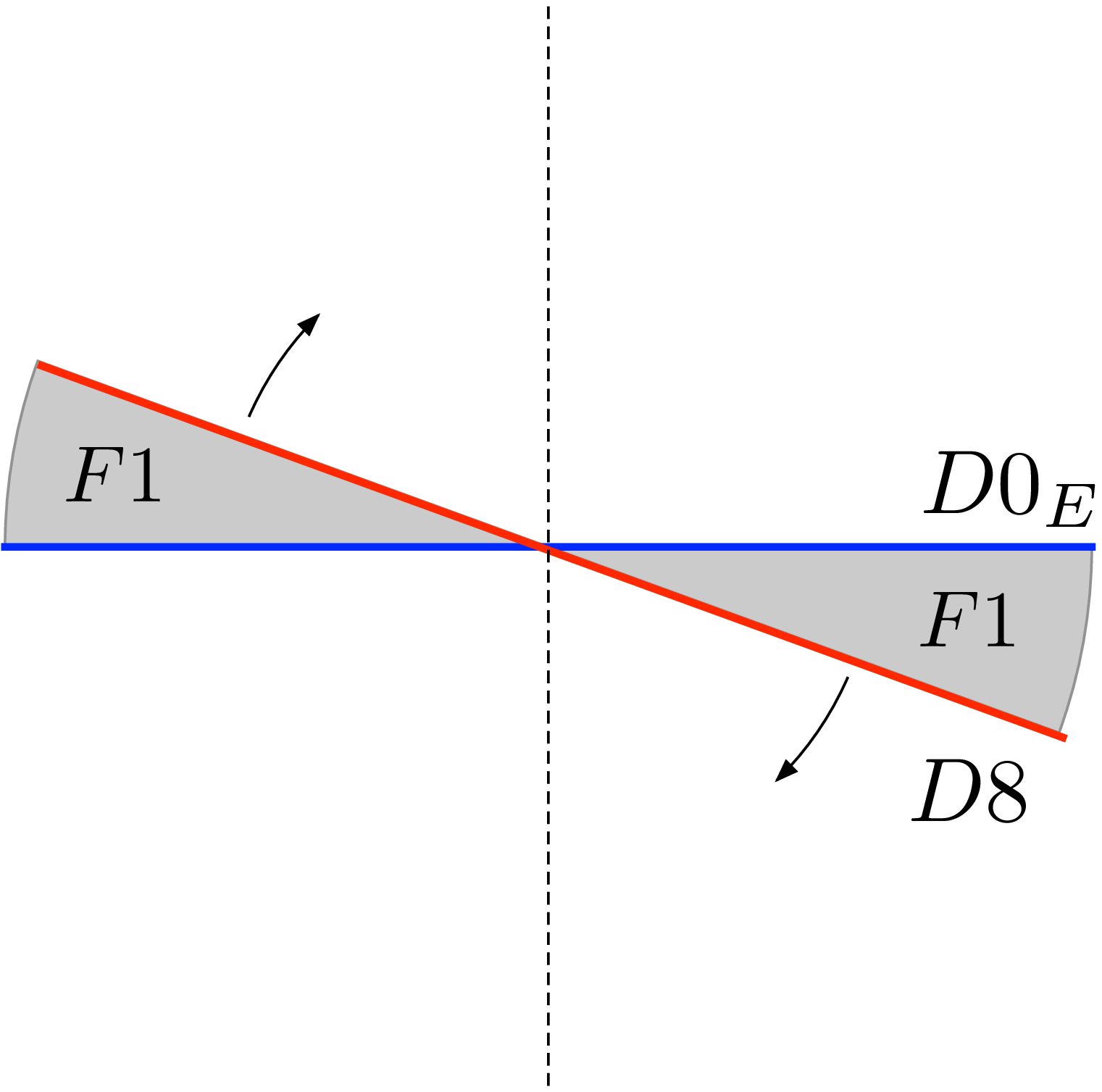,width=3.5cm}\hspace{0.5cm}
\epsfig{file=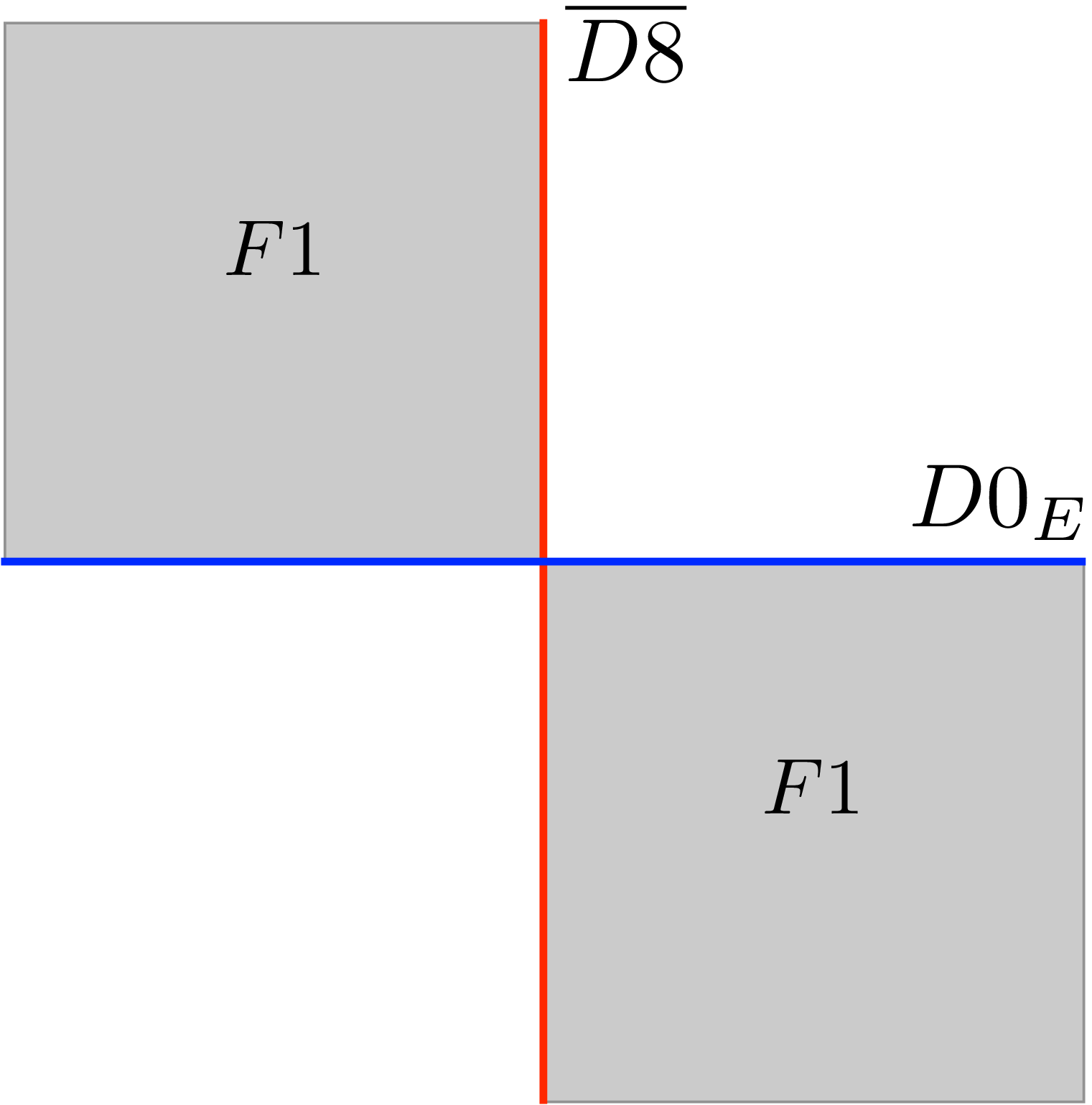,width=3.5cm}}
\caption{The creation of a Euclidean string worldsheet in a D$0_E$-D8 system.}
\end{figure}

\end{document}